\documentclass[12pt]{iopart}
\usepackage{graphicx}
\usepackage{bm}
\usepackage{iopams} 
\usepackage{datetime}
\usepackage{etoolbox}

\begin{document}
\title{Prospects for single-molecule electrostatic detection in molecular motor gliding motility assays}

\author{M.~Sanchez Miranda$^{1}$, R.~Lyttleton$^{1,2}$, P.H.~Siu$^{1}$, S.~Diez$^{3}$, H.~Linke$^{2}$, A.P.~Micolich$^{1}$}
\address{$^{1}$ School of Physics, University of New South Wales, Sydney NSW
2052, Australia}
\address{$^{2}$ NanoLund and Solid State Physics, Lund University, S-22100 Lund, Sweden}
\address{$^{3}$ B CUBE - Center for Molecular Bioengineering, Technische Universit\"{a}t Dresden, D-01307 Dresden, Germany}
\ead{adam.micolich@nanoelectronics.physics.unsw.edu.au}
\submitto{\NJP}
\date{\today}

\begin{abstract}
Molecular motor gliding motility assays based on myosin/actin or kinesin/microtubules are of interest for nanotechnology applications ranging from cargo-trafficking in lab-on-a-chip devices to novel biocomputation strategies. Prototype systems are typically monitored by expensive and bulky fluorescence microscopy systems and the development of integrated, direct electric detection of single filaments would strongly benefit applications and scale-up. We present estimates for the viability of such a detector by calculating the electrostatic potential change generated at a carbon nanotube transistor by a motile actin filament or microtubule under realistic gliding assay conditions. We combine this with detection limits based on previous state-of-the-art experiments using carbon nanotube transistors to detect catalysis by a bound lysozyme molecule and melting of a bound short-strand DNA molecule. Our results show that detection should be possible for both actin and microtubules using existing low ionic strength buffers given good device design, e.g., by raising the transistor slightly above the guiding channel floor. We perform studies as a function of buffer ionic strength, height of the transistor above the guiding channel floor, presence/absence of the casein surface passivation layer for microtubule assays and the linear charge density of the actin filaments/microtubules. We show that detection of microtubules is a more likely prospect given their smaller height of travel above the surface, higher negative charge density and the casein passivation, and may possibly be achieved with the nanoscale transistor sitting directly on the guiding channel floor.
\end{abstract}

\maketitle

\section{Introduction}

A key frontier in nanotechnology in recent decades has been the development of functional nanodevices featuring biomolecular motors, with a strong focus on motile `gliding' assays using the actin-myosin and kinesin-microtubule molecular motor systems~\cite{vandenHeuvelSci07, GoelNatNano08, AgarwalProgPolySci10, HessARBE11}. Gliding assays were initially developed to better understand the underpinning biophysics of the actin-myosin~\cite{KronPNAS86, HaradaCMCS88} and kinesin-microtubule~\cite{HowardNat89} motor systems. Directed motion of actin filaments~\cite{SuzukiBPJ97, NicolauBPJ99} and microtubules~\cite{HiratsukaBPJ01} via lithographically-patterned guiding structures opened the path towards motile biosensors~\cite{RamachandranSmall06} and associated diagnostics~\cite{KortenCurrOpBiotech10}, nanoscale cargo-trafficking systems~\cite{BrunnerLoC07, GoelNatNano08, SchmidtLoC10}, bio-inspired optical devices~\cite{AoyamaPNAS13}, and most recently, structures for efficiently solving combinatorial problems by having molecular motors explore a complex network~\cite{NicolauPNAS16}.

Network-based biocomputation~\cite{NicolauPNAS16} is a strong motivator to extend beyond microscope-based detection of filaments because scale-up inevitably leads to the need to detect filament passage at a very large number of points spread across an area much larger than the field-of-view for an appropriate magnification objective. This need for new detection strategies led us to consider an electrical approach where a nanoscale transistor made with, e.g., carbon nanotubes, graphene or semiconductor nanowires, spans the bottom of a lithographically-patterned guiding structure. Both actin filaments~\cite{TuszynskiBPJ04} and microtubules~\cite{vandenHeuvelPNAS07} carry negative charge that can in principle electrostatically gate a nanoscale transistor providing it passes within a distance smaller than the buffer's Debye length. Such an electrostatic sensor for gliding molecular motor assays could have more widespread applications, for example, as a trigger for optical~\cite{HessNL01} or thermal~\cite{KortenNL12} actuation of motility, direction of traffic at junctions~\cite{vandenHeuvelSci06}, addition/removal of cargo~\cite{GoelNatNano08, HessARBE11}, or addition/removal/read-out of chemical `tags' added to filaments to improve information carrying capacity~\cite{MicolichArXiv19}.

Our goal here is to establish the feasibility and develop some design guidelines for such a nanoscale transistor detector. We do this via basic `first principles' calculations accounting for key aspects of the sensor geometry and design, combined with detection estimates based on published experimental data for detection of other biological targets using nanoscale transistor devices. Our calculations show that electrostatic detection of passing filaments is certainly possible in principle. We define and address key experimental parameters, e.g., buffer ionic strength, protein charge, etc., and discuss how they influence the limits of detection for this type of sensor. Notably, we find that electrostatic detection is likely to be a significantly easier prospect for microtubules than it is for actin.

\section{Background and Approach}
The development of nanoscale transistors featuring carbon nanotube, graphene or semiconductor nanowire conducting channels for biosensing applications has a long history~\cite{ZhangChemRev16}. We have chosen to focus on a transistor featuring one single-walled carbon nanotube as the transistor channel for two main reasons. First, the low diameter ($1-2$~nm) provides a strongly confined but highly exposed conduction channel that enables strong gate response~\cite{ZhangChemRev16}. Semiconductor nanowires are much larger ($10-100$~nm diameter) with less strongly confined carriers that are set slightly back from the surface due to band-bending and surface oxidation effects~\cite{UllahNanotech17}. Graphene falls between the two but involves an additional free parameter, channel width, which adds complexity to the calculations. Second, there are numerous experimental works characterizing the response of single-walled carbon nanotube transistors to single biomolecules~\cite{GoldsmithNL08, SorgenfreiNatNano11, ChoiSci12} and aspects related to Debye screening effects in these devices~\cite{SorgenfreiNL11, ChoiNL13}. These help make our task considerably easier than it might otherwise be.

\begin{figure}
\begin{center}
\includegraphics[width = \textwidth]{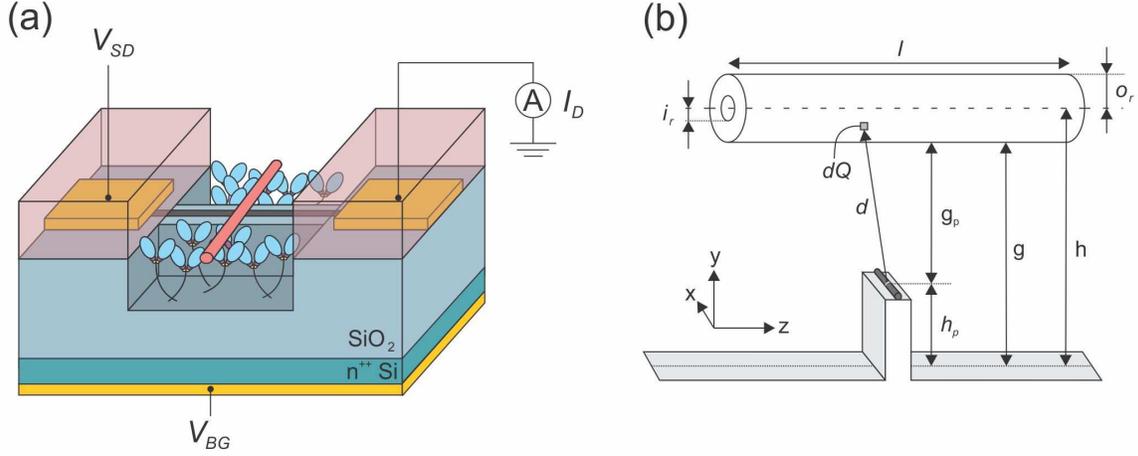}
\vspace{5mm}
\end{center}
\caption{(a) Schematic of the detector concept, which consists of a carbon nanotube transistor integrated into the guiding channel structure typically used in gliding motility assays (not to scale). The $n^{+}$-Si substrate (green) is used as a back-gate for the carbon nanotube transistor, which is connected to source and drain contacts (gold) buried under a material (pink) used to define the guiding channel. A voltage $V_{SD}$ applied to the source contact drives a current $I_{D}$ through the nanotube, which is measured flowing out of the drain contact. The transistor is tuned to its most sensitive condition, i.e., highest $dI_{D}/dV_{BG}$, via the back-gate potential $V_{BG}$ applied to a metal contact to the $n^{+}$-Si substrate. The carbon nanotube and its source and drain contacts are insulated from the back-gate by a SiO$_{2}$ layer (light blue), which can also be used to raise the nanotube above the floor of the guiding channel to overcome Debye screening (see text). (b) Schematic of the electrostatic model, which features a filament with inner radius $i_{r}$, outer radius $o_{r}$, length $l$ and charge per length $\lambda$, travelling along the $z$-direction with its centre of mass at a height $h$ above the guiding channel floor. The free-space between the bottom of the filament and the nanotube $g = h - o_{r}$ is referred to as the elevation following Ref.~\cite{KerssemakersPNAS06}. A carbon nanotube at $z = 0$ runs along the $x$-direction on an oxide pedestal with height $h_{p}$ to reduce the effective filament elevation at the detection point to $g_{p}$. The carbon nanotube is treated as a one-dimensional conductor. We integrate over infinitesimal charge elements $dQ$ located a distance $d$ away from the most strongly gated segment of the nanotube at $(0, 0, 0)$ to obtain the electrostatic potential difference $\Delta V$ generated by a passing filament.}
\end{figure}

\subsection{Detector Concept}
Figure~1 shows a schematic of our detector concept. A carbon nanotube transistor spans a microfabricated channel for guiding actin filaments or microtubules, which we hereafter simply call `filaments' unless there is a need to be specific to one or the other. The key challenge for an electrostatic sensor is getting the transistor close enough to the filament to overcome Debye screening. Filaments travel some finite distance above the guiding channel surface, lifted by the surface-bound motors that provide their motive force. The filaments have finite thickness meaning that there are two relevant heights -- the height of their centre-of-mass axis above the surface $h$ and the separation between their bottom edge and the surface $g$, which we refer to as the elevation. Ultimately $g = h - o_{r}$ where $o_{r}$ is the outer radius of the filament, i.e., half the thickness. For this work we use $h = 38$~nm ($g = 35.5$~nm) for actin~\cite{PerssonLangmuir10} and $h = 29.5$~nm ($g = 17$~nm) for microtubules~\cite{KerssemakersPNAS06}. The travel height is not easily tuned and is notably larger than the typical Debye length $\lambda_{D} \sim 0.5-5$~nm. This urges us to consider a mechanism for raising the carbon nanotube transistor by some height $h_{p} < 50$~nm relative to the guiding channel floor, as shown in Fig.~1, to give an effective sensor elevation $g_{p}$ that is significantly less than $g$.

A more detailed concept for the detector is outlined in Fig.~1(a) and is built on a degenerately-doped Si substrate covered with a $0.1-0.5~\mu$m layer of thermal oxide, as is typical for nanotube transistors~\cite{GoldsmithNL08, SorgenfreiNatNano11, ChoiSci12}. This SiO$_{2}$-on-Si structure gives a `back-gate' to which a voltage $V_{BG}$ can be applied for tuning the transistor operating point, i.e., the current $I_{D}$ that flows in response to an applied source-drain bias $V_{SD}$~\cite{ChoiSci12}, to maximise the sensitivity to changes in electrostatic potential local to the carbon nanotube. The $n^{+}$-Si can also be used as a mirror-surface for implementing fluorescence interference contrast (FLIC)~\cite{ParthasarathyCBCBP04} to enhance local optical detection~\cite{KerssemakersPNAS06, LardSciRep13}. The nanotube transistor's source and drain contacts and associated interconnects are buried under the material that defines the guiding channel structure to encapsulate them from the liquid environment inside the assay flow-cell. A realistic approach to fabricating such a structure is described in the Supplementary Information. With a basic device concept sketched out, we focus entirely for the remainder of this work on theoretical considerations, leaving aspects of experimental realisation, e.g., prototypes, for separate work.

The theoretical aspects of the approach can be roughly divided into two parts. First, we use a fairly simple electrostatic model, implemented in Python, to estimate the change in electrostatic potential produced at a nanotube transistor by a passing actin filament or microtubule. Second, we work from published experiments, in particular, Choi {\it et al.}~\cite{ChoiSci12} and Sorgenfrei {\it et al.}~\cite{SorgenfreiNatNano11} to establish the limits of detection for a change in electric potential at a nanotube transistor due to a biological target, e.g., lysozyme catalysis and DNA melting, respectively. These two parts are then combined to determine whether the change in electrostatic potential generated by a passing filament would be detectable.

\subsection{Electrostatic Model}
The electrostatic model is implemented in three dimensions~\cite{Note1} at the plane passing vertically through the filament perpendicular to the guiding channel floor, as illustrated in Fig.~1(b). We assume the carbon nanotube is a one-dimensional conductor (diameter $\rightarrow 0$) extending along the $x$-axis at $y = 0, z = 0$. The $x$-direction is included in the model simply to properly treat the screening for the cylindrical filaments. There is an implied geometrical assumption in this, which is that the guiding channel and the filament both run parallel to $z$ and are centred at $x = 0$. Ultimately, the only section of the nanotube we are interested in is the infinitesimal segment at $(0, 0, 0)$ where the strongest gating will occur by symmetry. In our work, we focus on knowing the effective gating potential at the nanotube due to the negatively-charged filament to enable direct empirical connection to prior experimental observations of biomolecular charge sensing in nanotube transistors~\cite{SorgenfreiNatNano11, ChoiSci12}. We have deliberately done this instead of modelling the conduction/response of the nanotube itself to ensure tighter connection to likely experimental reality. For simplicity, we also assume that the filament is fully rigid, i.e., has infinite persistence length, such that the travel height $h$ is the same at every point along its length. We thus ignore any bending of the filament caused either by the platform structure or myosin/kinesin binding. This could be an interesting extension for future work as height fluctuations due to filament flexibility could enhance the transient signal beyond our estimates.

We establish the potential $V$ at the nanotube segment at $(0, 0, 0)$ by considering the filament as a cylinder (solid for actin, hollow for microtubules) composed of equal point charges $dQ$. We integrate over the cylinder to account for the differing distances $d$ between the element $dQ$ and the nanotube segment at $(0, 0, 0)$. We designed the software for positional versatility along $z$, but if the centre of the filament is directly above $(0, 0, 0)$, we can speed up the calculation by integrating over only a quarter of the filament due to symmetries about $z = 0$ (left/right in Fig.~1(b)) and $x = 0$ (front/back in Fig.~1(b)). We obtain the gating potential difference $\Delta V$ as the difference between the potential $V$ with the filament directly overhead the nanotube (centred at $z = 0$) and with the filament at $z >> \lambda_{D}$, where we assume $V = 0$. This means that $\Delta V = 0 - V = -V$. For simplicity, we account for Debye screening using the Debye-H\"{u}ckel model,~\cite{DebyePhysZ23} noting that while we are not in a regime where counterion condensation~\cite{OosawaJPS57, ManningARPC72} can be ignored, we can reasonably account for it in our consideration of the effective filament charge density~\cite{ManningJCP69} (see \S2.2.2). By this approach we obtain a voltage contribution $dV$ for $dQ$ as:

\begin{equation} dV = \frac{dQ}{4\pi\epsilon_0\epsilon_r}\frac{1}{d}e^{-d/\lambda_D} \end{equation}

\noindent where $\epsilon_{r} \simeq 80$ for H$_{2}$O. Additionally, the Debye screening length $\lambda_D$ is~\cite{IsraelachviliBook11}:

\begin{equation} \lambda_D = \frac{1}{\sqrt{8\pi \lambda_{B} I N_{A}}} \end{equation}

\noindent where $N_{A}$ is Avogadro's number and ionic strength $I$ is in units of mM. The parameter $\lambda_{B}$ is the Bjerrum length, the separation where the electrostatic interaction between two charges is equal to $kT$:

\begin{equation} \lambda_{B} = \frac{e^{2}}{4\pi\epsilon_{0}\epsilon_{r}kT} \end{equation}

\noindent  Typically $\lambda_{B} = 6.95{\AA}$ at $27^{\circ}$C. The buffer ionic strength $I = \frac{1}{2} \Sigma \rho_{i}z_{i}^{2}$ where $\rho_{i}$ is the density of ion species $i$ and $z_{i}$ is its valence. We will return to consider all the buffers we used in our calculations below, as this is a crucial part of this problem and the viability of any electrostatic detection experiment.

A key component of the electrostatics is that the path from an element $dQ$ to the nanotube segment at $(0, 0, 0)$ in Fig.~1(b)is rarely entirely through buffer alone. Indeed, for an element in the upper half of a microtubule, the direct path can pass from protein to buffer, back to protein (lower half of microtubule), and then through buffer the rest of the way to the nanotube (see, e.g., Fig.~S4). A reader's initial perception might be that this should not be a major problem. After all, a common first year undergraduate electrostatics problem is a capacitor with a multi-layer dielectric. The capacitor problem is tractable because there are a pair of finite equipotentials with finite separation that provide suitable boundary conditions. However, for the integration of point charges that we do here, a change in material means that the boundary conditions, namely $V = \infty$ at $r = 0$ and $V = 0$ at $r = \infty$, are no longer finite. This makes the change of material analytically intractable.~\cite{Note2} That said, what we seek here is an estimate rather than an exact result and there are two aspects that play to our favour -- both are reliant on the protein largely displacing the buffer in its vicinity. The first is that the dielectric constant for protein $\epsilon_{p}$ is less than the dielectric constant for water $\epsilon_{w}$. By assuming $\epsilon_{w}$ over the entire path irrespective of material, our resulting potential $V$ at the nanotube will always be an under-estimate because the dielectric constant appears in the denominator in Eqn.~1. In other words, the inability to deal with the dielectric change means that any signal we measure should be somewhat larger than what we calculate and so, in terms of determining detection threshold, gives us a significant margin of safety. The second aspect is that Debye screening, as an exponential term, is usually dominant. Thus the dielectric change is a secondary factor to the estimate as long as the path length through the buffer is known so that the screening contribution is accurate. This means a key component of our software is the geometric mapping of the transition points between protein and buffer along the path between each $dQ$ and the nanotube -- for actin there is only ever one but for microtubules there are often three. This need for accuracy in path length components is another factor that motivated our approach for this problem. We elaborate on the geometry and algorithm used for this part of the analysis in the Supplementary Information (see Figs.~S2-S4). The assumption that displacement of the buffer by the protein limits screening is justified further in our discussion on the studies made by Choi {\it et al.}~\cite{ChoiSci12, ChoiNL13} and the detection limit estimates.

A nice aspect of our situation above is that we now do not need to account for the dielectric constant of the protein $\epsilon_{p}$ in the software at least; we do use it for the detection limit estimates. This is fortunate because the magnitude of $\epsilon_{p}$ is a matter of significant debate with suggested values that vary depending on the specific protein, its structure and even the depth within it being considered. Ultimately, this means that $\epsilon_{p}$ may not even be constant over the path segment inside the filament in our problem. We have assumed $\epsilon_{p} = 5$ as a compromise position from several references~\cite{PiteraBJP01, SchutzProtein01, LiJCTC13}. A final point regarding the modelling. More sophisticated 3D models of the overall system and device might seem appealing, for example, electrostatics by finite element analysis~\cite{ChuahNatComm19, LevyJColIntSci20} combined with transistor modelling by non-equilibrium Green's function approaches~\cite{NeophytouAPL06}. However, uncertainties in various aspects ranging from $\epsilon_{p}$ and buffers to device layout and geometry might outweigh any gains arising from the increased sophistication these models would provide, at least initially.

\subsubsection{Buffer Solutions}
We consider a set of nine buffer solutions here, three for actomyosin assays and six for kinesin-microtubule assays. The respective buffers, their ionic strength and corresponding Debye length are presented in Table~1 (see also Fig.~S5). The incentive from a detection perspective is to maximise the Debye length to improve the chances of detection. This has a natural limit because the motility assays fail if the ionic strength becomes too low. We calculate buffer strength using the Henderson-Hasselbalch equation for the dominant buffer components; the values may differ slightly from some experimentally measured values~\cite{Note3} but the uncertainties here should not substantially change our conclusions. The buffers a60 and BRB-80 are commonly used buffers for actomyosin and kinesin, respectively, in the literature. They are the highest ionic strength buffers considered. The two additional actomyosin buffers (a40 and a20) give functional assays, whilst ionic strength much lower than a20 can result in rapidly declining motile velocity~\cite{HomsherAJP92, BalazAnalBiochem05} and/or depolymerisation~\cite{MaruyamaJBiochem84}. For kinesin, we consider a lower ionic strength BRB buffer (BRB-20) as well as three imidazole-buffers (I50Mg05, I50Mg05-2$\times$ and I50Mg05-3$\times$) and one HEPES buffer, all of which still give reasonably good motility in assays. More complete details of buffer ionic strength calculations and experimentally determined buffer ionic strengths are given in the Supplementary Information for completeness.

\begin{table}
    \begin{center}
        \begin{tabular}{|l||c|c|c|}
            \hline
            Buffer & $I$~(mM) & $\lambda_{D}$~(nm) \\
            \hline
            a60 & $51.06$ & $1.36$ \\
            BRB-80 & $106.76$ & $0.94$ \\
            \hline
            a40 & $31.06$ & $1.75$ \\
            a20 & $11.06$ & $2.93$ \\
            BRB-20 & $29.69$ & $1.79$ \\
            I50Mg05 & $16.14$ & $2.43$ \\
            I50Mg05-2$\times$ & $8.07$ & $3.43$ \\
            I50Mg05-3$\times$ & $5.38$ & $4.20$ \\
            HEPES & $6.23$ & $3.90$ \\
            \hline
        \end{tabular}
    \end{center}
\caption{Ionic strength $I$ and Debye length $\lambda_{D}$ for the eight buffers considered in this paper. The actomyosin assay buffers a60, a40 and a20 are $10$~mM MOPS at pH~$7.4$ with $1$~mM MgCl$_{2}$ and $5$, $25$ or $45$~mM KCl~\cite{NicolauPNAS16, tenSiethoffPhD13}. The kinesin assay buffers BRB-80 and BRB-20 are $80$~mM PIPES and $20$~mM PIPES at pH~$6.8$ with $1$~mM KCl and $1$~mM EGTA~\cite{NicolauPNAS16}. The I50Mg05, I50Mg05-2$\times$ and I50Mg05-3$\times$ buffers are $50$~mM Imidazole buffers at pH~$6.8$ with $0.5$~mM MgCl$_{2}$ used direct, diluted $1:1$ and diluted $1:2$ in water, respectively~\cite{ReutherPC18}. The HEPES buffer is $5$~mM HEPES at pH~$7.2$ with $0.25$~mM EGTA, and $0.25$~mM MgCl$_{2}$~\cite{ReutherPC18}.}
\end{table}

\subsubsection{Filament parameters}
The remaining parameters in the model all relate to the filaments themselves. Table~2 highlights the key parameters and the values used with literature sources. The inner radius $i_{r}$ and outer radius $o_{r}$ are straightforward. The filament length $l$ is taken as $1~\mu$m for all calculations, noting that the filament length is physically insignificant providing $l >> \lambda_{D}$. We included $l$ in the code simply to test for physically realistic outcomes in key limits, and to enable future consideration of short filaments in very low ionic strength buffers.

The challenging parameter is the charge per length $\lambda$ for two reasons: a) we need to account for counterion condensation as mentioned earlier, and b) literature values can vary significantly. Actin filaments and microtubules, like DNA, are linear polyions. Counterion condensation occurs when the charge per unit length $\lambda$ becomes sufficient that the charge spacing $l_{charge} \sim 1/\lambda$ exceeds the Bjerrum length $\lambda_{B}$, leading to a phase transition where a fraction $1 - 1/\Gamma$ (with $\gamma = \lambda_{B}/l_{charge}$) of the counterions `condense' to form a layer of thickness $R_{M}$ (the Manning radius) surrounding the polyion~\cite{OosawaJPS57, ManningARPC72}. This reduces the polyion's effective charge outside $R_{M}$ to less than the critical value, i.e., $l_{charge} < \lambda{B}$~\cite{OosawaJPS57,ManningARPC72}. The exact fraction condensed and hence the effective charge of a polyion relative to its bare charge depends on details of the buffer solution, but an effective rough approximation is to assume the effective charge is half of the bare charge~\cite{KalraACSNano20}. The crucial point here is that even when counterion condensation occurs, the Debye-H\"{u}ckel model~\cite{DebyePhysZ23} remains reasonably valid outside the counterion condensation layer as long as the effective charge accounting for counterion screening is used~\cite{ManningJCP69}. This enables our model to remain valid as long as the filament-sensor separation $g_{p}$ exceeds a few Bjerrum lengths (a few nm). In the instance where $g_{p}$ does become small, the sensor sees more of the bare charge, so our model likely becomes an under-estimate rather than an over-estimate of signal. The critical charge for counterion condensation onto filaments is of order $1400$~e/$\mu$m based on a Bjerrum length of $6.95{\AA}$, meaning counterion condensation cannot be ignored for either actin or microtubules based on the charge density values that follow.

Turning back to linear charge density values for our model, we will deal with microtubules first as the literature is more substantive. The bare charge for tubulin is $48$~e/dimer~\cite{NogalesNat98} corresponding to a linear charge density of $84,375$~e/$\mu$m assuming $1750$~dimers per $1~\mu$m of microtubule~\cite{JiaBMD04}. The effective charge density is commonly measured experimentally by electrophoresis with reported values ranging from $\sim 0.5$~e/$\mu$m~\cite{JiaBMD04} to $\sim 270$~e/$\mu$m~\cite{StrackeBBRC02, KimBPJ08} to $>30,000$~e/$\mu$m~\cite{MinouraBPJ06, vandenHeuvelPNAS07}. van den Heuvel {\it et al.}~\cite{vandenHeuvelPNAS07} explain the lower values as arising from assuming only hydrodynamic friction and ignoring the retarding effect of the counterions. Notably, their measured value of $37,400$~e/$\mu$m corresponds to $23$~e/dimer, which is very close to $50\%$ of the bare charge~\cite{vandenHeuvelPNAS07}. We thus take this as our default linear charge density for microtubules for this work.

Regarding the bare charge for actin, Tang \& Janmey~\cite{TangJBC96} have each actin monomer as carrying $14$ negative charges in vacuum, reducing to $11$ per subunit on protonation of His residues in H$_{2}$O, which gives $4070$~e/$\mu$m assuming $370$ monomers per $\mu$m. This value is consistent with the $4000$~e/$\mu$m given as a theoretical value in vacuum by Tuszynski {\it et al.}~\cite{TuszynskiBPJ04}. Electrophoresis studies on actin are limited and none of them give a value for the charge per length. However, Tang \& Janmey~\cite{TangJBC96} demonstrate counterion screening at $40-80\%$ level, depending on buffer conditions via actin bundle formation studies, suggesting that an effective charge density of $2000$~e/$\mu$m should be a reasonable default value for this work. For completeness, we note a value of $16,500$~e/$\mu$m given by Cantiello {\it et al.}~\cite{CantielloBPJ91}, which appears at odds with the values above. We include this value in our range when we study the influence of changing linear charge density in Fig.~4 and in the Supplementary Information.

\begin{table}
    \begin{center}
        \begin{tabular}{|l||c|c|c|}
            \hline
            Parameter & Actin & Microtubules \\
            \hline
            $i_{r}$ & $0$~nm~\cite{ChasanEBPJ02} & $8.4$~nm~\cite{SchaapBPJ06, vandenHeuvelPNAS07} \\
            $o_{r}$ & $2.5$~nm~\cite{ChasanEBPJ02} & $12.5$~nm~\cite{SchaapBPJ06, vandenHeuvelPNAS07} \\
            $h$ & $38$~nm~\cite{PerssonLangmuir10} & $29.5$~nm~\cite{KerssemakersPNAS06} \\
            $g$ & $35.5$~nm~\cite{PerssonLangmuir10} & $17$~nm~\cite{KerssemakersPNAS06} \\
            $\lambda$ & $37,400$~e/$\mu$m~\cite{vandenHeuvelPNAS07} & $2,000$~e/$\mu$m \\
            \hline
        \end{tabular}
    \end{center}
\caption{Default filament parameters used in the calculations unless otherwise specified. The parameters are inner radius $i_{r}$, outer radius $o_{r}$, free height $h$, elevation $g$ and charge per length $\lambda$.}
\end{table}

\subsection{Detection Limit Estimate}
Our detection limit estimates are based on studies of a single lysozyme molecule bound to a single-wall carbon nanotube transistor by Choi {\it et al.}~\cite{ChoiSci12, ChoiNL13}. We also made estimates based on ssDNA melting studies using point-defect functionalized single-wall carbon nanotube transistors by Sorgenfrei {\it et al.}~\cite{SorgenfreiNatNano11, SorgenfreiNL11} for completeness; these are presented in the Supplementary Information. Choi {\it et al.} study a T4 lysozyme molecule bound to the nanotube via a pyrene-maleimide linker molecule attached to the C90 residue~\cite{ChoiSci12, ChoiNL13}. Their buffer solution is $10$~mM sodium phosphate buffer at pH~$7.5$ with $10-300$~mM NaCl. The corresponding Debye length ranges from $1.64$~nm at $10$~mM NaCl to $0.54$~nm at $300$~mM NaCl with $\lambda_{D} = 1.12$~nm at the $50$~mM NaCl concentration used in most of the work (full details in the Supplementary Information). The key to our estimate is the ability to establish the charged residue producing the signal. Choi {\it et al.}~\cite{ChoiNL13} identify these as the K83 and R119 residues based on sequence/structure considerations and studies using targeted mutation. Both residues are $\sim 1.5$~nm away from the C90 residue where the few {\AA}-long molecule binding the protein to the nanotube is attached. We assume that the strongest gating occurs at the carbon nanotube attachment site. This is partly due to proximity and partly because the $\pi-\pi$ stacking at this site should withdraw some carriers from conduction, increasing the local resistance and taking the attachment site vicinity closer to full depletion~\cite{ZhaoAdvMater08, ChoiSci12}. For purposes of this paper, we assume the attachment point in Choi {\it et al.}~\cite{ChoiSci12, ChoiNL13} corresponds to $(0, 0, 0)$ in our model.

\begin{figure}
\begin{center}
\includegraphics[width = \textwidth]{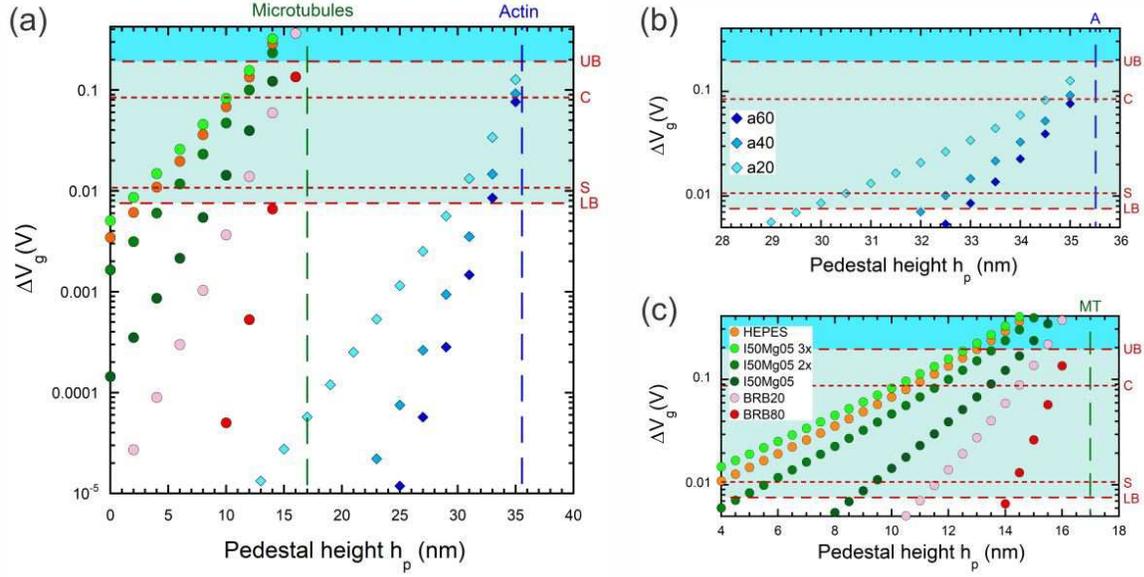}
\vspace{5mm}
\end{center}
\caption{Plots of voltage step $\Delta V_{g}$ vs pedestal height $h_{p}$ for actin (diamonds) and microtubules (circles) for the nine buffers in Table~1. We focus on the full $h_{p}$ range in (a) with a finer focus on the detection range for actin in (b) and microtubules in (c). The dashed vertical lines indicate the natural elevation for microtubules (green - $g = 17$~nm; denoted MT in (b)) and actin (blue - $g = 35.5$~nm; denoted A in (c)). The horizontal dashed lines indicate detection limits and are labelled UB = upper bound ($192$~mV), LB = lower bound ($7.6$~mV), C = Choi~\cite{ChoiNL13} experimental result ($85$~mV) and S = Sorgenfrei~\cite{SorgenfreiNatNano11} experimental result ($10.7$~mV). Instances in the green shaded region between LB and UB are likely to be detectable, instances in the blue shaded region above UB are highly likely to be detectable. Note that we ignore the surface casein layer~\cite{HowardMCB93} used in microtubule assays in this calculation; this is dealt with in Fig.~3.}
\end{figure}

Lysozyme is a processive enzyme that catalyses cleavage of peptidoglycan substrate. The active site for catalysis is $3.1$~nm from C90 and thus should be outside direct detection range according to Choi {\it et al.}~\cite{ChoiNL13}. We do not disagree, but simply note there is currently no definitive experimental evidence that solidly confirms or denies this position. Instead, Choi {\it et al.}~\cite{ChoiNL13} suggest catalysis causes $0.15$~nm movements for the K83 and R119 residues relative to the attachment site~\cite{KurokiPNAS99, ChoiNL13}, which is the asserted mechanism for their electrostatic detection. Note that these motions can be detected individually for each residue via targeted mutation, and are still detected at [NaCl]$~= 300$~mM where $\lambda_{D} = 0.54$~nm~\cite{ChoiNL13}. We expect this ability to see charge displacement at a range well beyond $\lambda_{D}$ occurs because the protein displaces the buffer solution between the K83 and R119 residues and the C90 attachment point. This reduces both the screening and the intervening dielectric constant. It also motivates our comment about not being able to definitively rule out seeing some small effect from the catalytic site itself.

Considering this from a device context, the carbon nanotube transistor used by Choi {\it et al.}~\cite{ChoiSci12, ChoiNL13} experiences a shift in the location of nearby charged residues as an effective gate voltage $V_{g}$, which causes a shift in the average measured drain current $I_{D}$ as a function of time. The measured transconductance $dI_{D}/dV_{BG}$ local to the operating point, obtained from a separate measurement of $I_{D}$ vs back-gate voltage $V_{BG}$ (e.g., Fig.~1C in Ref.~\cite{ChoiSci12}), can be used to convert a change in average current $\Delta I_{D}$ into an effective gate voltage change $\Delta V_{g} = \Delta I_{D}(dI_{D}/dV_{BG})^{-1}$. The typical transconductance can vary considerably between devices ranging from $\sim 25$~nA/V in Ref.~\cite{ChoiSci12} to $\sim 100$~nA/V in Ref.~\cite{ChoiNL13}. Choi {\it et al.}~\cite{ChoiNL13} obtain the effective gate voltage change $\Delta V_{g}$ that results by changing the net charge of the K83 and R119 residues by targeted mutation using the average current and transconductance. In doing so they obtain $\Delta V_{g} = 84 \pm 2$~mV/$e$. Considering this from the perspective of a simple point-charge model is interesting. The targeted mutation is equivalent electrostatically to adding/removing a charge $e$ situated $1.5$~nm from the C90 attachment point, which can be estimated as:

\begin{equation} \Delta V_{G} = \frac{e}{4\pi\epsilon_{0}\epsilon_{r}}\frac{1}{d}e^{-d/\lambda_{D}} \end{equation}

If we do this assuming $\epsilon_{r} = 80$ (buffer) and $\lambda_{D} = 0.54$~nm, we obtain $\Delta V_{g} = 745~\mu$V. In contrast, if we assume $\epsilon_{r} = 5$ (protein) and $\lambda_{D} \rightarrow \infty$, we obtain $\Delta V_{g} = 192$~mV. If we further assume that a small amount of screening occurs over the length of the maleimide moiety in the pyrene-maleimide linker, a distance of order $4.5$~{\AA} with $\lambda_{D} = 0.54$~nm, then we obtain $\Delta V_{g} = 83$~mV. This is very close to the experimental value obtained by Choi {\it et al.}~\cite{ChoiNL13}. It is a notable outcome because it leads to two conclusions. First, it confirms that screening only really occurs in the tiny gap between C90 and pyrene attachment to the nanotube. This supports our earlier assertion that the direct line of sight from K83 and R119 to C90 runs through the body of the protein, where water and ions are largely excluded~\cite{Note4}. Second, it suggests that a simple electrostatic treatment can give realistic order-of-magnitude estimates for real experimental signal.

We now apply this analysis to the scenario in Ref.~\cite{ChoiSci12} to obtain a `detection limit' under the premise that this is one of the most sensitive carbon nanotube sensors experimentally realised under conditions similar to those we would face in a motility assay. To achieve this we calculate the change in potential that results for a singly charged residue (either K83 or R119 but not both) moving by $0.15$~nm at a distance of $1.5$~nm from the C90 attachment point. If we assume $\epsilon_{r} = 5$ over the entire distance, since it is mostly filled with protein, this gives us an upper bound on detection limit of $\Delta V_{g} = 192$~mV. We can also include a small screening contribution with $\lambda_{D} = 0.54$~nm for the $4.5$~{\AA} maleimide moiety (see above), which gives a lower bound on detection limit of $\Delta V_{g} = 7.6$~mV. Note that these span the experimental estimate of $85 \pm 2$~mV given by Choi {\it et al.}~\cite{ChoiNL13} We refer to upper and lower bounds here because the most accurate answer would involve including the screening but having $\epsilon_{r} = 80$ and not $5$ over the $4.5$~{\AA} linker molecule gap. However, dealing with this $\epsilon_{r}$ transition at the protein edge is analytically intractable (see earlier). Physically, we know the increased $\epsilon_{r}$ would act to slightly lessen the screening, giving a result somewhere between the bounds of $192$~mV and $7.6$~mV that we calculated above. Since our focus here is on mathematical simplicity, we will continue just with the upper and lower bound estimates and the experimental value from Choi {\it et al.}~\cite{ChoiNL13} in our results.

Finally, to corroborate these values, we also performed an estimate based on the work by Sorgenfrei {\it et al.}~\cite{SorgenfreiNatNano11, SorgenfreiNL11}. Full details are given in the Supplementary information, but we estimate a detection limit of $10.7$~mV for their device, which agrees well with their own estimates of $5-11$~mV~\cite{SorgenfreiNL11}.

\section{Results}

From this point onwards we focus entirely on the output of our electrostatic modelling for actin filaments and microtubules. Figure~2 shows plots of effective transient gate voltage step $\Delta V_{g}$ that we expect a passing filament to produce versus the height $h_{p}$ of the pedestal that raises the carbon nanotube above the guiding channel floor. Figure~2(a) covers the full range of $h_{p}$ for both filament types for a broader comparative picture. Figures~2(b/c) present the same calculations with finer $h_{p}$ resolution and tighter focus on the possible detection regime for actin (Fig.~2(b)) and microtubules (Fig.~2(c)). Data for actin is presented as diamonds and microtubules as circles with colour assigned to corresponding buffer composition. The horizontal dashed lines indicate detection limits with labelling as designated in the figure caption. The background shading indicates our expectation of the likelihood of electrostatic detection being viable, and ranges from unlikely (white) to likely (green) and highly likely (blue). The vertical dashed lines indicate the natural elevations for microtubules (green, marked MT) and actin (blue, marked A). We limit our analysis to $h_{p} < g$ as a natural outcome of our assumption of infinite persistence length (all parts of the filament are always at elevation $g$).

\begin{figure}
\begin{center}
\includegraphics[width = 0.5\textwidth]{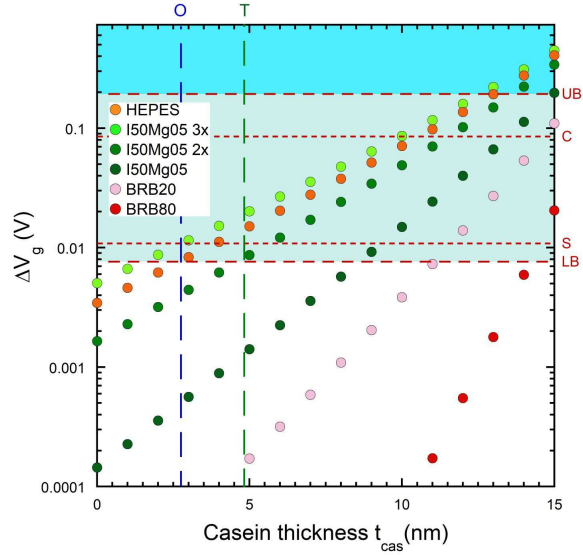}
\vspace{5mm}
\end{center}
\caption{Voltage step $\Delta V_{g}$ vs casein layer thickness $t_{cas}$ for pedestal height $h_{p} = 0$ for microtubules with the six relevant buffers in Table~1; there is no corresponding layer in an actin motility assay. The dashed vertical lines indicate the effective fully-screening casein thickness corresponding to the measured details of $\beta$-casein-on-SiO$_{2}$ layers obtained by Tiberg {\it et al.}~\cite{TibergBMM01} using neutron reflectometry (green and denoted T; $t_{cas} = 4.8$~nm) and Ozeki {\it et al.}~\cite{OzekiBPJ09} (blue and denoted O; $t_{cas} = 2.75$~nm) using quartz crystal microbalance studies. We extend the $t_{cas}$ axis to $15$~nm to cover the maximal thickness given by Verma {\it et al.}~\cite{VermaJBiolEng08}, which is a very optimistic upper-bound as discussed in the text. The horizontal dashed lines indicate detection limits and are labelled UB = upper bound, LB = lower bound, C = Choi~\cite{ChoiNL13} experimental result and S = Sorgenfrei~\cite{SorgenfreiNatNano11} experimental result. Instances in the green shaded region between LB and UB are likely to be detectable, instances in the blue shaded region above UB are highly likely to be detectable.}
\end{figure}

Looking at Fig.~2(a), there are two obvious behaviours in the data. The first is that the signal $\Delta V_{g}$ strengthens as the carbon nanotube is raised closer to the filament. This is a natural consequence of reduced distance $d$ between the charge elements $dQ$ and the point of action (0, 0, 0) via Eqn.~1. The second is that $\Delta V_{g}$ at a given $h_{p}$ increases as the buffer ionic strength is reduced, arising from the $\lambda_{D}$ contribution to Eqn.~1. The slope $\Delta V_{g}/h_{p}$ also decreases with reduced ionic strength, however, slope is a deceptive concept here due to the logarithmic axis for $\Delta V_{g}$. Note that even with the logarithmic axis, the $\Delta V_{g}$ versus $h_{p}$ is not truly linear in these plots because there is a $1/d^{2}$ Coulomb term in addition to the exponential Debye screening contribution, i.e., the data curves gently even with the log axis for $\Delta V_{g}$. Care needs to be taken with linear extrapolation for data presented as we have it in Fig.~2.

Regarding detection prospects, we start by considering actin filaments and return to microtubules afterwards. The data in Fig.~2 makes it clear that detection of actin is impossible without the nanotube on a pedestal of at least $h_{p} = 30$~nm. Improvement in detection prospects can be achieved by reducing buffer ionic strength -- the stronger improvement in moving from a40 to a20 compared to moving from a60 to a40 in Fig.~2(b) arises from $\lambda_{D} \sim 1/\sqrt{I}$, as is visually evident in Fig.~S5. However, there is a limit to reducing ionic strength due to the biochemical reasons outlined earlier. Thus there is no escaping the need for a pedestal, and the associated fabrication challenges in achieving it, for detecting passing actin filaments at the charge density expected based on the literature~\cite{TangJBC96, TuszynskiBPJ04}. We explore the effects of increasing the filament charge density later in the paper.

Turning to microtubules, there are three notable aspects compared to actin that are vital to the discussion. First, microtubules travel closer to the guiding channel floor, i.e., have lower $g$, which means our scope for possible detection without a pedestal structure, i.e., with $h_{p} = 0$, is substantially improved. Second, the charge density for microtubules is an order of magnitude higher. Thirdly, an important aspect of microtubule assays that we have so far ignored is the use of a casein coating on the substrate~\cite{HowardMCB93}. The casein layer serves to anchor the kinesin tail and position the kinesin head to better interact with microtubules~\cite{VermaJBiolEng08, OzekiBiophysJ09}. Actin assays have no corresponding layer; the myosin tails bind directly to the silanized SiO$_{2}$ surface~\cite{tenSiethoffPhD13}. The data in Fig.~2 ignores the casein layer, enabling us to make a more direct comparison with actin purely on the basis of elevation, charge density and buffer ionic strength. We will add the casein layer for microtubules to our model momentarily (Fig.~3).

The data in Fig.~2 with a bare SiO$_{2}$ surface for both filament types also points to the need to raise the nanotube on a pedestal for microtubules, albeit to a much lesser extent. For the commonly used PIPES buffers BRB-80 and BRB-20, a pedestal with $h_{p} = 13-16$~nm would be required. In contrast, the imidazole and HEPES buffer results, particularly for I50Mg05-3$\times$ and HEPES are such that one could potentially imagine detection occurring without a pedestal at all, e.g., due to fluctuations arising from finite persistence length. Thus before we even consider the casein layer, microtubules have an obvious advantage from a device fabrication perspective owing to charge density and buffer ionic strength range alone.

\begin{figure}
\begin{center}
\includegraphics[width = \textwidth]{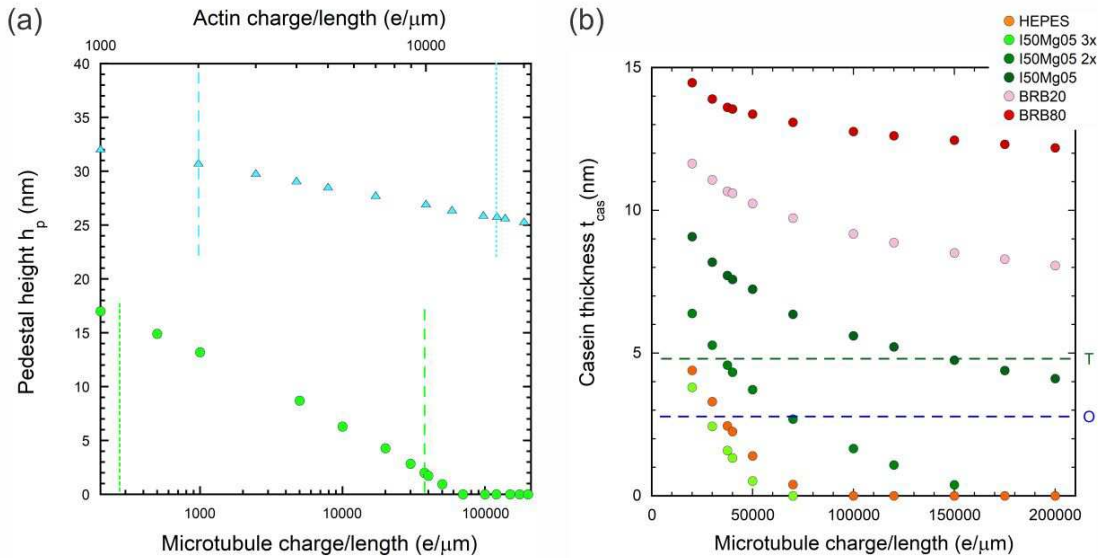}
\vspace{5mm}
\end{center}
\caption{(a) Pedestal height $h_{p}$ required to achieve a voltage step $\Delta V_{g}$ exceeding the lower bound for detection (LB in Fig.~2(a)) as a function of filament charge density $\lambda$ for actin (a20 buffer, blue diamonds, top axis) and microtubules (I50Mg05-3$\times$ buffer and no casein layer, green circles, bottom axis). The vertical dashed lines indicate the default charge density for actin (blue - $2000$~e/$\mu$m) and microtubules (green - $37,400$~e/$\mu$m). The vertical dotted lines indicate other notable charge density values with the actin value obtained by Cantiello {\it et al.}~\cite{CantielloBPJ91} (blue - $16,500$~e/$\mu$m) and the microtubule value obtained by Stracke {\it et al.}~\cite{StrackeBBRC02} and Kim {\it et al.}~\cite{KimBPJ08} (green - $270$~e/$\mu$m). (b) Casein thickness $t_{cas}$ required to achieve a voltage step $\Delta V_{g}$ exceeding the lower bound for detection (LB in Fig.~3) with no pedestal $h_{p} = 0$ for microtubules for the six associated buffers in Table~1. The dashed horizontal lines indicate the effective fully-screening casein thickness corresponding to the measured details of $\beta$-casein-on-SiO$_{2}$ layers obtained by Tiberg {\it et al.}~\cite{TibergBMM01} using neutron reflectometry (green and denoted T; $t_{cas} = 4.8$~nm) and Ozeki {\it et al.}~\cite{OzekiBPJ09} (blue and denoted O; $t_{cas} = 2.75$~nm) using quartz crystal microbalance studies.}
\end{figure}

Adding the casein layer to our model is relatively straightforward and we use an approach similar to how we deal with exclusion of the buffer by the body of the filament itself. We consider the casein layer to have thickness $t_{cas}$, and use geometric arguments (see Fig.~S3/4) to work out what this corresponds to as a path length $d_{cas}$ on the direct line from $dQ$ to the point of action $(0, 0, 0)$. We then switch the screening term off for that segment of the path, as we do with the segment passing through the body of the filament. Note that this results in a slight underestimate of signal because we cannot account for the change in dielectric strength for reasons described earlier. The greater challenge is working out a meaningful value to assign for $t_{cas}$. Whole casein is a mixture of four casein sub-groups -- $\alpha_{s1}$, $\alpha_{s2}$, $\beta$ and $\kappa$ -- in proportions that vary based on species of origin and processing~\cite{EigelJDairySci84}. The composition of the surface-adsorbed casein layer for a microtubule assay can vary and has a significant effect on microtubule binding~\cite{VermaJBiolEng08} and motility~\cite{MaloneyPLoSOne11}, yet is rarely precisely specified/known for many assays in the literature (e.g., often non-specific descriptors like `casein-containing buffer' appear in literature protocols). Of the four caseins, only $\beta$-casein is well characterized in terms of formation of a layer on solid surfaces. Our view of the literature is that there is general agreement that $\beta$-casein forms a bilayer on a hydrophilic surface such as SiO$_{2}$. $\beta$-casein is amphiphilic~\cite{EigelJDairySci84} and the bilayer forms as a two-step process: a) first the hydrophilic domain adsorbs to the SiO$_{2}$ surface giving a tightly-packed monolayer presenting a hydrophobic outer surface, then b) a second loosely-packed monolayer forms as hydrophobic domains adsorb to the monolayer presenting a hydrophilic outer surface for the final bilayer~\cite{VermaJBiolEng08, NylanderLangmuir97, TibergBMM01, OzekiBPJ09}. The disagreement in the literature is more related to the thickness and packing density for the $\beta$-casein bilayer. Verma {\it et al.}~\cite{VermaJBiolEng08} point to Nylander {\it et al.}~\cite{NylanderLangmuir97} as evidence for $t_{cas} = 15$~nm. However, that measurement is obtained by compressive force measurements between a pair of $\beta$-casein bilayers on opposing hydrophilic surfaces~\cite{NylanderLangmuir97}. As such, it detects the maximal steric extent of the bilayer, and given the looser packing of the outer monolayer and its hydrophilicity, assuming full suppression of screening over $15$~nm is probably an over-estimate. We thus consider this an absolute upper-bound on our calculations, setting the plotted range in Fig.~3. Most papers on $\beta$-casein layers suggest $t_{cas}$ for our estimate should be smaller. Tiberg {\it et al.}~\cite{TibergBMM01} report neutron reflectometry data for $\beta$-casein on SiO$_{2}$ and suggest the bilayer is best modelled as a trilayer with a) $3$~nm thickness at $63\%$ volume fraction, b) $3.5$~nm thickness at $51\%$ volume fraction and c) $4.0$~nm thickness at $28\%$ volume fraction. We treat this as an effective fully non-screening layer of thickness $4.8$~nm for simplicity. Ozeki {\it et al.}~\cite{OzekiBPJ09} report quartz crystal microbalance (QCM) measurements for $\beta$-casein on a SiO$_{2}$-functionalized QCM crystal and suggest an inner layer of $2.1$~nm thickness with a density of $279$~ng/cm$^{2}$ and an outer layer that is either $1/3$ of the thickness at the same density or $1/3$ of the density at the same thickness. As an effective single non-screening layer the corresponding thickness would be $2.75$~nm.

In Figure~3, we plot $\Delta V_{g}$ versus the casein thickness $t_{cas}$ (fully non-screening) assuming a pedestal height $h_{p} = 0$. In other words, we focus on the question of how thick the casein layer needs to be to achieve detection without needing a pedestal at all. We run the model out to $t_{cas} = 15$~nm to allow Verma {\it et al.}~\cite{VermaJBiolEng08} as a very optimistic upper-bound. We indicate the more realistic scenarios in Ozeki {\it et al.}~\cite{OzekiBPJ09} (effective $t_{cas} = 2.75$~nm) and Tiberg {\it et al.}~\cite{TibergBMM01} (effective $t_{cas} = 4.8$~nm) with blue and green dashed vertical lines respectively. The $\Delta V_{g}$ for the I50Mg05-3$\times$ buffer is very close to the lower-bound detection limit at the $t_{cas} \sim 2.5-5$~nm range corresponding to the realistic effective $t_{cas}$ thickness~\cite{TibergBMM01, OzekiBPJ09} that one might expect using purely $\beta$-casein. The HEPES and I50Mg05-2$\times$ buffer results are also reasonably close to the detection limit in this range.

However, $\beta$-casein is only a small part of a larger picture. We focus on it in Fig.~3 because it is the only casein sub-group that is well characterised as an adsorbed film on SiO$_{2}$ in terms of thickness and volume fraction. Maloney {\it et al.}~\cite{MaloneyPLoSOne11} performed studies of microtubule motility in assays using pure $\beta$-casein surface passivation and found a) few motile filaments, b) that the filaments that were motile tended to be quite long, and c) weaker attachment at the ends of the microtubule, which led to poor tracking and a short time before microtubules detached and floated away. They instead found that whole casein ($49\%$ $\alpha$, $37\%$ $\beta$, $14\%$ $\kappa$) was a much better option, supporting stable motility for a wide range of filament lengths with good filament velocity and reproducibility, with pure $\alpha$-casein performing nearly as well~\cite{MaloneyPLoSOne11}. The thickness and volume fraction for $\alpha$-casein and whole casein passivation layers on SiO$_{2}$ is not presently available, but if they increase the effective thickness of the casein layer from a buffer displacement perspective without increasing the filament elevation, i.e., move us further to the right of the two scenarios in Fig.~3, then the prospects of electrostatic detection of microtubules with $h_{p} = 0$ only improve. Given this, we suggest a characterisation of whole casein and $\alpha$-casein on SiO$_{2}$ using neutron reflectometry or QCM studies would be a worthwhile endeavour.

Before moving on, we make three final observations regarding casein passivation. First, $\beta$-casein has a net negative charge of $-7.5~e$/molecule at pH $6-7$~\cite{Perez-FuentesMat17}; the other caseins are likely also charged. This charge would cause an offset in the transistor's maximum sensitivity point (i.e., peak transconductance), which would need to be corrected by applying a back-gate voltage $V_{BG}$. It is presently unclear whether the required $V_{BG}$ would: a) be within the available range for nanotube transistors, as too high a $V_{BG}$ leads to dielectric breakdown of the SiO$_{2}$ layer, destroying the device, or b) is compatible with retaining casein adsorption, kinesin binding and microtubule motility. The former could be established simply by comparing nanotube transistors before/after casein passivation, without the need for other structures, e.g., guiding channels, or a full motility assay. The latter is readily testable without the carbon nanotube device structure. Both would be worthwhile experiments in their own right. We note that relatively strong transverse electric fields ($10-100$~kV/m) were not a significant issue for microtubule gliding assays~\cite{vandenHeuvelSci06}, so there may well not be major problems with the vertical electric fields associated with gate-tuning of the nanotube transistor either. Second, carbon nanotubes are generally~\cite{Note5} hydrophobic~\cite{KyakunoJCP16} and casein will bind with its hydrophilic domains down against the SiO$_{2}$ substrate. This means there is some chance the casein layer is affected/interrupted local to the nanotube, e.g., incomplete, thinner than usual, etc. We would not immediately expect this to affect motility, but it might mean that the screening is slightly stronger than expected based on Fig.~3. Some focussed studies might also be required to establish this in an experimental context. Third, one could ask why a similar strategy could not be used to improve the prospects for actomyosin assays. A critical feature of why casein works in the microtubule assay is that the kinesin tails bind to the lower half of the casein bilayer with the kinesin heads positioned just above the casein bilayer surface~\cite{VermaJBiolEng08, OzekiBiophysJ09}. The crucial aspect for this work is that the addition of this layer does not increase filament elevation $g$; simply putting in an underlayer that coats the nanotube but at the same time increases $g$ will not have the same effect in terms of reduced screening. We know of no such layer presently available for actomyosin assays.

As a last consideration in this paper, we turn to the possibility of varying the charge of the filaments used in motility assays. This could be due to uncertainty in the filament charge or it could be part of a deliberate charge-enhancement strategy, e.g., by binding short-strand DNA to the outside of a microtubule~\cite{IsozakiSciRep15, IsozakiSciRob17}. Figure~4(a) shows a plot of the pedestal height required to reach the detection limit, i.e., $\Delta V_{g}$ at the lower-bound (LB), versus the linear charge density $\lambda$ on the filament in $e/\mu$m. The data presented is for the lowest ionic strength buffers, a20 for actin and I50Mg05-3$\times$ for microtubules, in the latter case without the casein layer. The default charge densities are indicated by vertical dashed lines (blue for actin/green for microtubules) with two notable values from earlier discussion, namely $16,500$~e/$\mu$m for actin~\cite{CantielloBPJ91} and $270$~e/$\mu$m for microtubules~\cite{StrackeBBRC02, KimBPJ08}, indicated by vertical dotted lines as points of note (see also Fig.~S6/S7 for underlying data). For actin, increasing the charge only really reduces $h_{p}$. Rather extreme amounts of added charge would be needed to eliminate the pedestal entirely. In contrast, for microtubules, only a $33\%$ increase in $\lambda$ is required to achieve detection with $h_{p} = 0$ before we even consider the casein layer. We add the casein layer to our consideration in Fig.~4(b)and focus on the $t_{cas}$ required to achieve the detection limit with $h_{p} = 0$ for all six buffers used for microtubule assays. Decreasing the ionic strength too much can lead to poor motility, so knowledge of the highest ionic strength buffer viable for detection is useful. Our data in Fig.~3(b) already showed that microtubules should be observable for the I50Mg05-3$\times$ buffer at $h_{p} = 0$ and the data in Fig.~4(b) confirms that the charge density would need to be less than $1/3$ of that measured by van den Heuvel {\it et al.}~\cite{vandenHeuvelPNAS07} to not be detectable. A similar situation holds for the HEPES and I50Mg05-2$\times$ buffers, however increasing the charge density would certainly help. For the undiluted I50Mg05 buffer, $\lambda$ would need to increase by a factor of $5-10$ but might be within reasonable possibility. In contrast the PIPES buffers, BRB-80 and BRB-20 are still a long way from detectability even with large increases in $\lambda$ unless a commensurate improvement in the density/thickness of the casein layer can be obtained somehow. Ultimately, the data in Figs.~3 and 4 convinces us that the prospects for electrostatic detection using microtubules are strong with the carbon nanotube on the guiding channel floor ($h_{p} = 0$) providing lower ionic strength buffers than those commonly used for microtubule assays, i.e., BRB-80 and BRB-20, are deployed. The prospects can be further improved if the effective thickness/density of the casein layer can be improved and the microtubule charge can be increased slightly using an approach similar to that demonstrated by Isozaki {\it et al.}~\cite{IsozakiSciRep15, IsozakiSciRob17} and further work in those two directions is also encouraged.

Finally, we note two aspects that we have not accounted for in our calculations that might improve the detection prospects slightly for both assay types. The first is the finite persistence length of the filaments and thermal fluctuations. These lead to vertical movement of the filament segments between attachments to surface-bound motors. This effect can be enhanced by reducing the bound motor density on the substrate~\cite{VanDelinderSciRep19}. The second is that we have neglected the nanotube diameter, which at $1-2$~nm is already $25-50\%$ of the small pedestal height ($h_{p} = 4$~nm) required for microtubules when the casein layer was not taken into account. It would slightly improve detection prospects beyond our estimates for $h_{p} = 0$ when the casein layer is accounted for. This second contribution will be much less substantial for actin.

\section{Conclusions}

We have addressed the prospects for electrostatic detection of passing actin filaments and microtubules in a molecular motor motility assay by combining a basic electrostatic model for determining the effective gate potential at a carbon nanotube transistor due to a filament overhead with realistic detection scenarios from recent state-of-the-art nanotube transistor sensors~\cite{SorgenfreiNatNano11, ChoiSci12}. On a bare consideration of magnitudes, we suggest that both actin and microtubules are likely detectable electrostatically in a device context with sufficient effort. Actin requires the carbon nanotube to be raised on a pedestal of height $28-33$~nm above the guiding channel floor depending on buffer ionic strength; a possible fabrication route for achieving this is given in the Supplementary Information. Microtubules carry three distinct advantages that make them a much better prospect for electrostatic detection: a) higher charge density, b) lower natural travel elevation, and c) the presence of a casein passivation layer~\cite{HowardMCB93}, which incorporates/supports the kinesin and partially displaces buffer without changing the microtubule elevation. In contrast, for an actomyosin assay, the myosin molecules are bound directly to a silanized-SiO$_{2}$ surface with no equivalent to the casein passivation layer~\cite{tenSiethoffPhD13}. We suggest microtubules are likely detectable without the nanotube on a pedestal for the lowest ionic strength buffers when the casein passivation layer is accounted for. Another aspect that might improve detection prospects is increasing the charge density of the filament. Surface charge density modification for microtubules using short-strand DNA has already been demonstrated~\cite{IsozakiSciRep15, IsozakiSciRob17} and might be fruitful towards electrostatic detection.

Our results suggest that attempts at electrostatic detection would have a greater likelihood of success if focussed on kinesin/microtubules. Low ionic strength buffers, e.g., imidazole or HEPES, should be used rather than higher ionic strength PIPES buffers, e.g., BRB-80, and the casein layer should be prepared to optimise its thickness and volume fraction as much as possible to reduce screening contributions. Some obvious experiments that our work here also points to are a more detailed characterisation of the structural aspects, e.g., thickness and volume fraction, of $\alpha$-casein and whole casein passivation layers on SiO$_{2}$ to facilitate comparison with $\beta$-casein-on-SiO$_{2}$, which is better characterised~\cite{TibergBMM01, OzekiBPJ09} but less suited for motility assay performance~\cite{MaloneyPLoSOne11}. Studies of how casein passivation shifts the gate threshold voltage in carbon nanotube transistors and how motility depends on the back-gate voltage for an assay structure with an incorporated back-gate, e.g., SiO$_{2}$-on-$n^{+}$-Si, would also be useful towards further development of devices for electrostatic detection of filaments in gliding motility assays.

\section{Acknowledgements}

This work was funded by the Australian Research Council (ARC) under DP170104024 and DP210102085, the Volkswagen Foundation and the European Union's Horizon 2020 Programme under grant agreement No. 732482 (Bio4Comp). A.P.M. was a Japan Society for the Promotion of Science (JSPS) Long-term Invitational Fellow during the drafting of this manuscript. The process development work reported in the Supplementary Information was performed using the NSW node of the Australian National Fabrication Facility (ANFF). We gratefully acknowledge helpful discussions with Paul Curmi, Mercy Lard, Lawrence Lee, Frida Lindberg and Cordula Reuther in the course of this work.

\end{document}